\documentclass[12pt]{article}
\usepackage{amssymb,amsmath,latexsym,enumerate,graphicx,epsfig,cite,subfigure,indentfirst}
\numberwithin{equation}{section}
\begin{document}
\title{Power and power-logarithmic expansions for travelling-wave solutions of the Burgers-Huxley equation.}
\author{ Olga Yu. Efimova \and  Nikolai A. Kudryashov \and  Mikhail A. Chmykhov}
\date{}
\maketitle
\begin{abstract}
The Burgers-Huxley equation is studied. All power and
power-logarithmic expansions for travelling-wave solutions of this
equation are presented. Using the power expansions, some exact
solutions of this equation are found.
\end{abstract}
\section{Introduction.}
 The  Burgers-Huxley equation takes the form
\begin{equation}\label{e:B-H_eq}
u_t+\alpha u u_x=Du_{xx}+\beta u+\gamma u^2-\delta u^3, \qquad
D>0,\,\delta>0
\end{equation}
It is used for description of non-linear wave processes in physics,
ecology and economics \cite{Osipov01,Loskutkov01,Broadbridge01,
Bazyikin01, Svirizhev01}. The condition of positiveness for
coefficients $D$ and $\delta$ follows from the physical meanings of
the problems.

The  Burgers-Huxley equation is not the exactly solvable one.
However some exact solutions can be obtained
\cite{Kudryashov01,Efimova01,Zaitcev01}, if we use the singular
manifold method
\cite{Weiss01,Kudryashov02,Kudryashov03,Kudryashov04,Chouldhary01}.

Using travelling waves $u(x,t)=w(z),\ z=x-C_0t$ in equation
\eqref{e:B-H_eq}, we have
\begin{equation}
Dw_{zz}-\alpha w w_z +C_0 w_z+\beta w+\gamma w^2-\delta w^3=0
\end{equation}

Taking $z=z'\sqrt{2D/\delta}$ into consideration, we obtain
\begin{equation}\label{e:eq}
w_{zz}-\alpha w w_{z} +C_0 w_{z}+\beta w+\gamma w^2-2 w^3=0
\end{equation}
where coefficients $\alpha$, $\beta$, $\gamma$ and $C_0$ are
constants. Primes in \eqref{e:eq} are omitted.

In the general case equation \eqref{e:eq} does not pass the
Painlev\'{e} test, so it is important to find all the asymptotic
forms and power expansions for the solutions of this equation. For
that we use the power geometry methods
\cite{Bruno01,Bruno02,Bruno03,Efimova02}.

The outline of this paper is as follows. In section 2 we consider
the general properties of equation \eqref{e:eq}. In section 3 the
power expansions, corresponding to the apexes of the carrier of
equation \eqref{e:eq}, are found. Sections 4-9 are devoted to the
power and power-logarithmic expansions, corresponding to the edges
of the carrier of equation \eqref{e:eq}. In section 10 the examples
of exact solutions are given.

\section{The carrier of equation \eqref{e:eq}.}

The carrier of equation \eqref{e:eq} consists of points
$Q_1=(-2,1)$, $Q_2=(0,3)$, $Q_3=(0,1)$, $Q_4=(-1,2)$, $Q_5=(0,2)$,
$Q_6=(-1,1)$. However some of these points can be absent, if
coefficients of equation \eqref{e:eq}, corresponding to them, are
equal to zero.

\begin{figure}[h] % было [p]
\center%
\subfigure[]{
 \epsfig{file=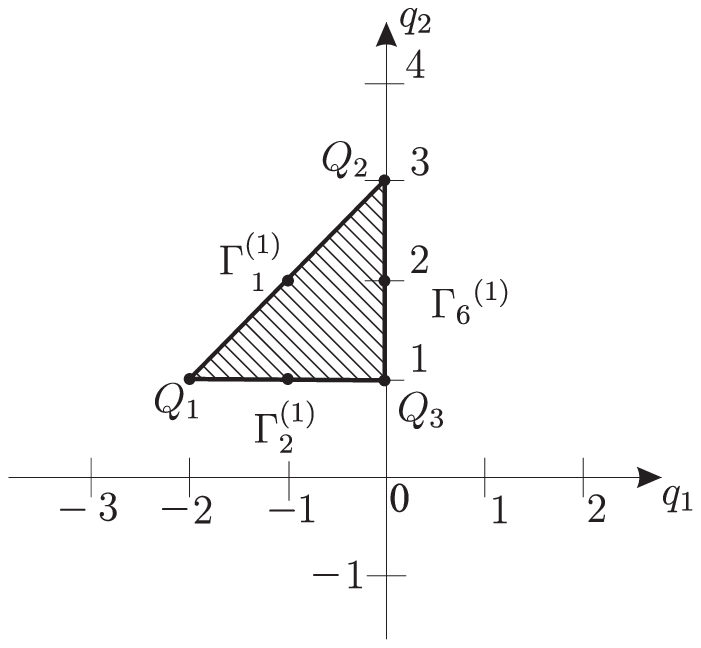,width=55mm}
\label{fig:f1a} } \subfigure[]{ \epsfig{file=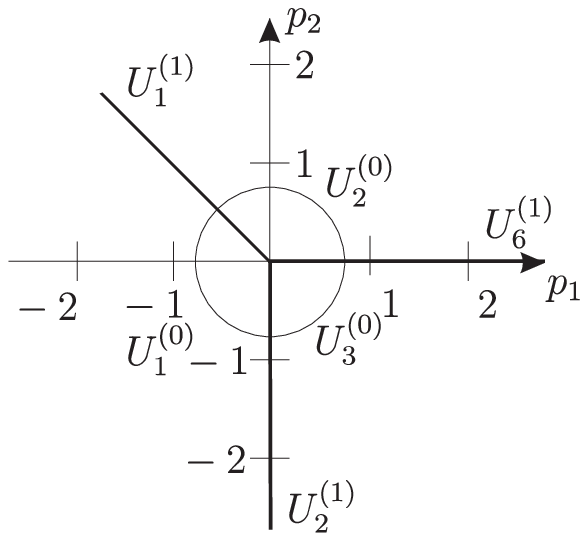,width=55mm}
\label{fig:f1b} } \caption{The carrier (a) and the normal cones
(b) of equation \eqref{e:eq} at $\beta\ne0$}\label{fig:f1}
\end{figure}

If $\beta\ne0$, the convex hull of the carrier of equation
\eqref{e:eq} is the triangle with apexes $Q_1$, $Q_2$, $Q_3$,
presented at fig. \ref{fig:f1}a.  The normal cones for apexes and
edges of the carrier of equation \eqref{e:eq} are given on fig.
\ref{fig:f1}b.

\begin{figure}[h] % было [p]
\center%
\subfigure[]{
 \epsfig{file=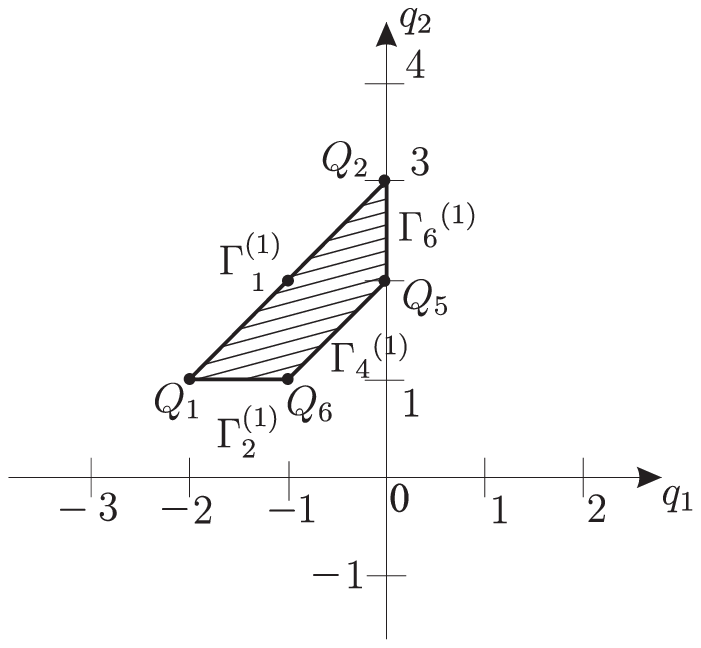,width=55mm}
\label{fig:f2a} } \subfigure[]{ \epsfig{file=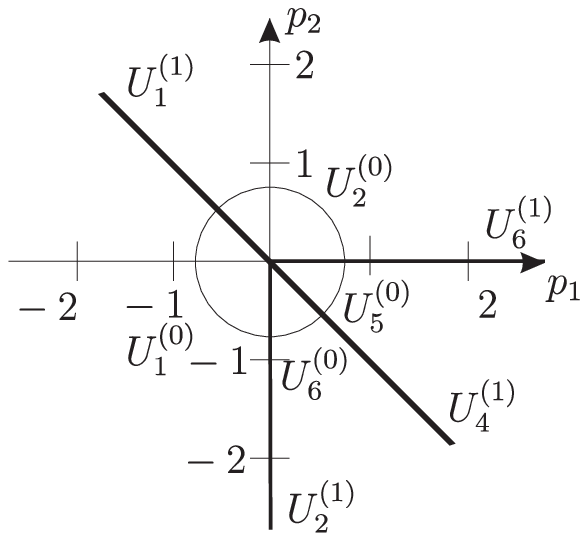,width=55mm}
\label{fig:f2b} } \caption{The carrier (a) and the normal cones
(b) of equation \eqref{e:eq} at $\beta=0$, $C_0\ne0$,
$\gamma\ne0$}\label{fig:f2}
\end{figure}

\begin{figure}[h] % было [p]
\center%
\subfigure[]{
 \epsfig{file=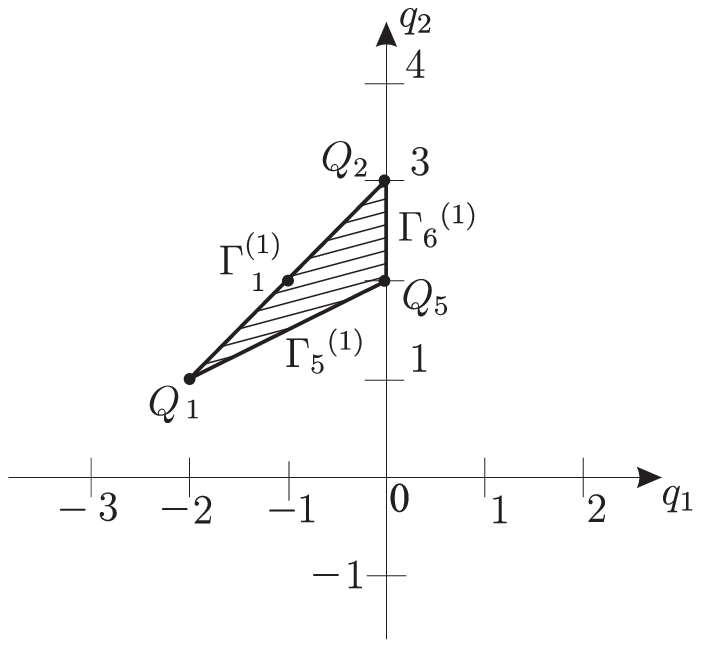,width=55mm}
\label{fig:f3a} } \subfigure[]{ \epsfig{file=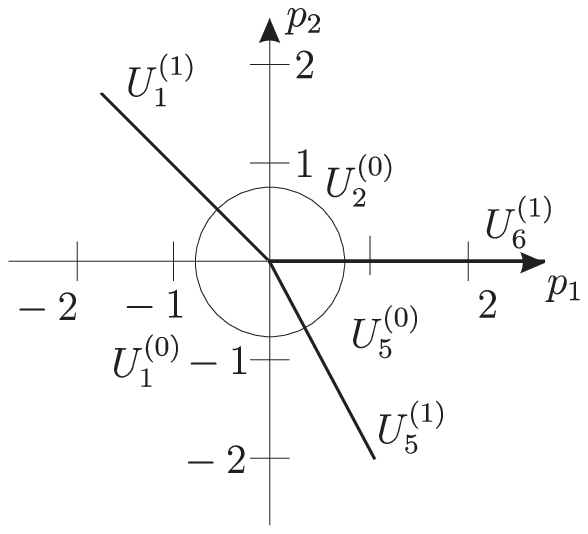,width=55mm}
\label{fig:f3b} } \caption{The carrier (a) and the normal cones
(b) of equation \eqref{e:eq} at $\beta=C_0=0$ and
$\gamma\ne0$}\label{fig:f3}
\end{figure}

\begin{figure}[h] % было [p]
\center%
\subfigure[]{
 \epsfig{file=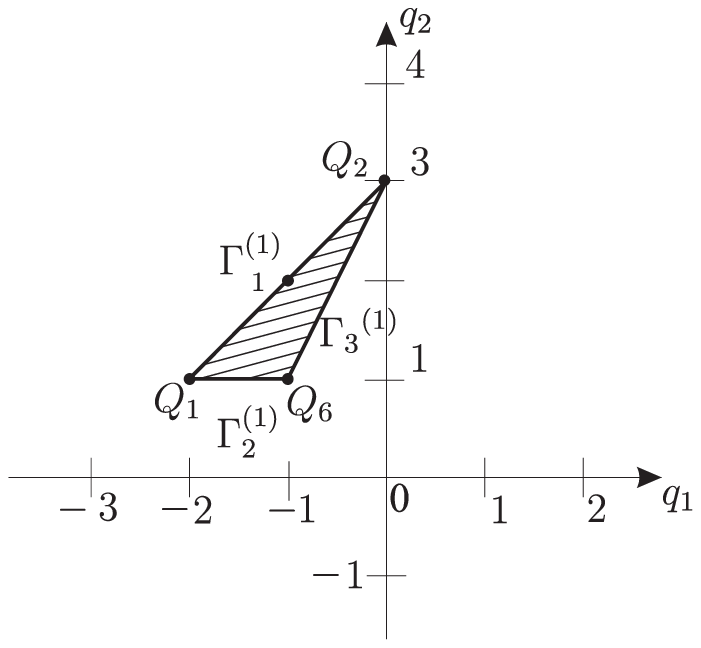,width=55mm}
\label{fig:f4a} } \subfigure[]{ \epsfig{file=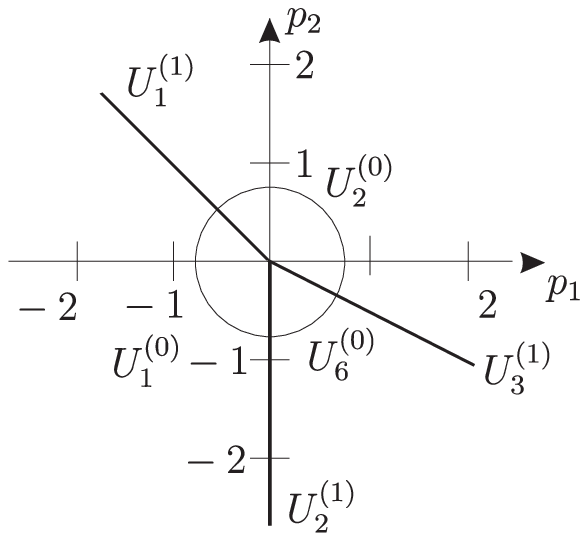,width=55mm}
\label{fig:f4b} } \caption{The carrier (a) and the normal cones
(b) of equation \eqref{e:eq} at $\beta=\gamma=0$ and
$C_0\ne0$}\label{fig:f4}
\end{figure}

In the case $\beta=0$, $\gamma\ne0$ and $C_0 \ne0$, the carrier of
equation \eqref{e:eq} is the trapezium, presented at fig. 2a.

If $\beta=0$ and one of the coefficients $\gamma$ or $C_0$ is
equal to zero, then the carrier and the normal cones of equation
\eqref{e:eq} are given on fig. 3 and fig. 4.

\section{Expansions, corresponding to the apexes of the carrier of equation \eqref{e:eq}.}
 The truncated equation, corresponding to apex $Q_1$ of the carrier of equation \eqref{e:eq}, takes the form
\begin{equation}
w_{zz}=0\label{e:uk_ur_v1}
\end{equation}
The characteristic polynomial for this equation is
\begin{equation} \chi(r)=r(1-r)
\end{equation}
It has roots $r=0$ and $r=1$.  Using the condition $\omega(1,r)\in
U_{1}^{(0)}$, we have
\begin{align*}r=0\Rightarrow \omega=-1,
z\rightarrow0\\r=1\Rightarrow \omega=-1, z\rightarrow0\end{align*}

For truncated equation  \eqref{e:uk_ur_v1} we obtain
$$\mathcal{L}(z)=\dfrac{d^2}{d z^2}\quad \Rightarrow  \nu(k)=k(k-1)$$ The proper numbers of the truncated equation
\eqref{e:uk_ur_v1} are $k=0$ and  $k=1$. The cone of the problem
is $\mathcal{K}=(k: k>r)$. So we have no critical numbers for
$r=1$ and one critical number for $k=1$.

The expansion of solutions, corresponding to the truncated
solution at $r=0$, takes the form
\begin{equation} w(z)= c_0+c_1 z+
\sum_{s=2}^{\infty}\,c_s\,z^{s}\label{e:v1_r=0}
\end{equation}
where $c_0$, $c_1$ are the arbitrary constants, other coefficients
can be sequentially calculated. Taking into account three terms we
obtain
\begin{equation}\label{e:p_ex_v_1} w(z)=c_{{0}}+c_{{1}}z+ \frac{1}{2}\left(
\alpha\,c_{{0}}c_{{1}}-\beta\,c_{{0 }}-\gamma\,{c_{{0}}}^{2}-C_0
c_{{1}}+2 {c_{{0}}}^{3} \right) {z}^ {2} +\ldots
\end{equation}

The expansion of solutions at $r=1$ is the special case of
expansion \eqref{e:v1_r=0} at $r=0$ with $c_0=0$. Taking into
account three terms we have the expansion for this case
\begin{equation}\label{e:p_ex_v_2}
w(z)=c_{{1}}z-\frac{1}{2}\,C_0 c_{{1}}{z}^{2}+\frac{1}{6}\,c_{{1}}
\left( -\beta+c_{{1}}\alpha +{C_0}^{2} \right) {z}^{3}+\ldots
\end{equation}

Depending on the parameters of equation  \eqref{e:eq}, its carrier
has two or three more apexes.

Consider the truncated equation, corresponding to apex
$Q_6=(-1,0)$. This apex exists if $\beta=0$ and
 $C_0\ne0$ (fig. 2, 4)
 \begin{equation}w_{z}=0\end{equation}
Here $r=0$. Condition $\omega(1,r)\in U_{6}^{(0)}$ does not hold,
so this apex does not give new expansions.

 The other truncated equations, corresponding to the apexes of the
 carrier of equation \eqref{e:eq}, are the algebraic one and,
 therefore, also have only trivial solutions.

%\section{Expansions, corresponding to the edges of the carrier of equation \eqref{e:eq}.}

\section{Expansions, corresponding to edge ${\Gamma_{1}}^{(1)}$.}

The truncated equation, corresponding to edge
${\Gamma_{1}}^{(1)}$, takes the form
\begin{equation}w_{zz}-\alpha w w_z - 2w^3=0 \label{e:uk_ur_r1}\end{equation}
The outer normal for this edge $N_{1}=(-1,1)$, so we have the
normal cone
$$U_{1}^{(1)}=\{\lambda(-1,1), \lambda>0\}$$ Using the condition $\omega(1,r)\in
U_{1}^{(1)}$, we obtain $r=-1$, $\omega=-1 \Rightarrow
z\rightarrow0$

The solution of truncated equation \eqref{e:uk_ur_r1} can be
described by formula
\begin{equation}\label{e:as_r1}w=\frac{c_{-1}}{z}\end{equation}
where \begin{equation}\label{e:c_for_r1}2{c_{-1}}^2-\alpha
c_{-1}-2=0
\end{equation} Equation \eqref{e:c_for_r1} has two solutions $$c_{-1}^{\pm}=\frac{\alpha}{4}\pm
\sqrt{\left(\frac{\alpha}{4}\right)^2+1}$$
 These roots correspond to two asymptotic forms. So it can be two
 power expansions.

For truncated equation \eqref{e:uk_ur_r1} we have
$$\mathcal{L}(z)=\dfrac{d^2}{d z^2}-\alpha \frac{c_{-1}}{z} \dfrac{d}{d z} + \frac{\alpha c_{-1}}{z^2}-6\frac{{c_{-1}}^2}{z^2}$$
Therefore the characteristic equation takes the form
$$\nu(k)=k(k-1)-\alpha c_{-1}k +\alpha c_{-1}-6{c_{-1}}^2$$
Using \eqref{e:c_for_r1}, we obtain the equation for proper
numbers
$$k^2-k(1+\alpha c_{-1}) -2\alpha
c_{-1}-6=0$$

This equation has roots $k_0=-2$ and
\begin{equation}k_1=3+\alpha c_{-1}\end{equation}

The cone of the problem is $\mathcal{K}=(k: k>-1)$, hence $k_0=-2$
is not the critical number.

Let us find, if power expansions exist under the different
meanings of the proper number $k_1$.

If $\alpha^2 \ne \dfrac{2a^2}{2+a}$, where $a\in Z, \ a>-2$, then
there are two power expansions in the form
\begin{equation}\label{e:exp_r1}w(z)=\frac{c_{-1}}{z}+\sum_{i=0}^{\infty}{c_{i}z^{i}}\end{equation}
where coefficient $c_{-1}$ is either root of equation
\eqref{e:c_for_r1}, other coefficients are sequentially computed.
In particular, $c_0$ and $c_1$ are determined by formulas
\begin{align*}&c_0= \frac {c_{-1}\gamma \alpha+2 \gamma-2 C_0
c_{-1}}{4(c_{-1} \alpha+3)}\\&c_1=\frac { c_{-1}( 2\gamma
c_0-6\,{c_0}^{2}+\beta
 ) }{3(c_{{-1}}\alpha+2)}
\end{align*}

%If $\gamma=C_0=0$, then in expansion \eqref{e:exp_r1} all
%coefficients with even indexes are equal to zero.

If $\alpha=0$ or $\alpha^2=2$, then $k_1$ is the critical number
for both expansions, and we should control the compatibility
conditions: at $\alpha=0$ to $c_3$ for both expansions; at
$\alpha^2=2$ to $c_5$ for one expansion and to $c_1$ for the other
one.

If $\alpha^2 = \dfrac{2a^2}{2+a}$ $(a=1,3,4,5,6,\ldots)$, then
$k_1$ is the critical number for only one of expansions, and we
should examine the compatibility condition to $c_{3+a}$. For the
other expansion all coefficients exist and can be uniquely
determined.

If the compatibility condition holds, then there is power
expansion with one arbitrary constant. This expansion has form, similar to expansion \eqref{e:exp_r1}.

If the compatibility
condition fails, then expansion is the power-logarithmic one and takes the form
\begin{equation}w(z)=\frac{c_{-1}}{z}+\sum_{i=0}^{\infty}{c_{i}(\ln z)z^{i}}\label{e:exp_r1_ln}\end{equation}
where coefficient $c_{-1}$ is either root of equation
\eqref{e:c_for_r1} and $c_{i}(\ln z)$ are the polinomials of $\ln z$ and can be uniquely determined.

For example let us consider $\alpha=0$ and $c_{-1}=1$ in details. In
this case we obtain the compatibility condition
\begin{equation}\label{e:cond_r1}C_0(\gamma+2C_0)(\gamma^2+\gamma C_0+9\beta-2{C_0}^2)=0 \end{equation}

If this condition holds, we have power expansion in the form
\begin{equation*}\begin{aligned} &w(z)=\frac{1}{z}+\frac{\gamma-C_0}{6}+\frac{z}{36}(6\beta-{C_0}^2+\gamma^2)
+\\&+z^2/216 (2C_0+\gamma)(\gamma^2+C_0\gamma+9\beta-2{C_0}^2))+c_3
z^3+\sum_{i=4}^{\infty}c_{i}z^{i}\end{aligned}\end{equation*} where  $c_3$ is the arbitrary constant
and $c_i,\,\,i=4,5,6...$ are the constants, that can be  sequentially defined.

If  the compatibility condition fails \eqref{e:cond_r1}, we obtain the
power-logarithmic expansion
\begin{equation*}\begin{aligned} w(z)=&\frac{1}{z}+\frac{\gamma-C_0}{6}+\frac{z}{36}(6\beta-{C_0}^2+\gamma^2)
+\\&+\frac{z^2}{216} (2C_0+\gamma)(\gamma^2+C_0\gamma+9\beta-2{C_0}^2))+\\&+
\left(c_3+C_0(2C_0+\gamma)(2{C_0}^2-C_0\gamma-\gamma^2-9\beta)\frac{\ln z}{270}\right)
z^3+\sum_{i=4}^{\infty}c_{i}(\ln z)z^{i}\end{aligned}\end{equation*} where $c_3$ is the arbitrary
constant and $c_i(\ln z)$ are the polynomials of $\ln z$, that can be uniquely
determined.

\section{Expansions, corresponding to edge ${\Gamma_{2}}^{(1)}$.}

Edge ${\Gamma_{2}}^{(1)}$ exists,  if $\beta\ne0$ or $C_0\ne0$.

 The
truncated equation, corresponding to edge ${\Gamma_{2}}^{(1)}$,
takes the form
\begin{equation}w_{zz}+C_0w_z+\beta w=0 \end{equation}

The outer normal $N_{2}=(0,-1)$, the normal cone
$$U_{2}^{(1)}=\{\lambda(0,-1),\lambda>0 \}$$

The condition $\omega(1,r)\in U_{2}^{(1)}$ does not hold, so there
are no power asymptotic forms, corresponding to this edge.

\section{Expansions, corresponding to edge  ${\Gamma_{3}}^{(1)}$.}

This edge exists, if $\beta=\gamma=0$ and $C_0\ne0$.

The truncated equation, corresponding to edge
${\Gamma_{3}}^{(1)}$, takes the form
\begin{equation}\label{e:uk_ur_r3}f w_{z}-2w^3=0\end{equation}

The outer normal for this edge is $N_{3}=(1,-1/2)$, therefore we
obtain the normal cone
$$U_{3}^{(1)}=\{\lambda (2,-1), \lambda>0\}$$
Using the condition $\omega(1,r)\in U_{3}^{(1)}$, we have
$\omega=1$, $r=-1/2 \Rightarrow z\rightarrow\infty$

So the solutions of truncated equation \eqref{e:uk_ur_r3} can be
written as
\begin{equation}w=\frac{c_0}{z^{1/2}}, \qquad \text{где}\,\,
c_0=\pm\sqrt{-C_0/4}
\end{equation}
The characteristic equation here is
$$\nu(k)=C_0(k+3/2)$$
We obtain one critical number $k=-3/2$. The compatibility condition
does not hold, so we obtain two power-logarithmic expansions
\begin{equation}\label{e:exp_r3}
\begin{aligned}
w(z) =&{\frac {c_{{0}}}{\sqrt {z}}}+{\frac {\alpha}{4 z}}+ \left(
c_{{2}}-{\frac {3c_{{0}}\ln z  }{{4C_0}}}
 \right) {z}^{-3/2}+\\&+ \left( {\frac {\alpha\, \left({\alpha}^{2} -32\,c_{{0}}
c_{{2}}-2 \right) }{16{C_0}}}-{\frac {3\alpha \ln z }{8C_0}} \right)
{z}^{-2}+\ldots
\end{aligned}
\end{equation}
where $c_0=\pm\sqrt{-C_0/4}$, $c_2$ is the arbitrary constant.

\section{Expansions, corresponding to edge  ${\Gamma_{4}}^{(1)}$.}

Edge ${\Gamma_{4}}^{(1)}$ exists, if $\beta=0$, $\gamma\ne0$ and
$C_0\ne0$.

The truncated equation, corresponding to edge
${\Gamma_{4}}^{(1)}$, takes the form
\begin{equation}\label{e:uk_ur_r4}C_0w_{z}+\gamma w^2=0\end{equation}

The outer normal for this edge $N_{4}=(1,-1)$, hence we have the
normal cone
$$U_{4}^{(1)}=\{\lambda (1,-1), \lambda>0\}$$

Using the condition $\omega(1,r)\in U_{4}^{(1)}$, we obtain
$\omega=1$, $r=-1 \Rightarrow z\rightarrow\infty$

So the solution of truncated equation can be written as
\begin{equation}u=\frac{C_0} {\gamma }z^{-1}
\end{equation}
For truncated equation \eqref{e:uk_ur_r4} we have
$$\mathcal{L}(z)=C_0\dfrac{d}{d z}+2\frac{C_0} {z^{-1}}\quad\Rightarrow
\nu(k)=C_0(k+2)$$

As the cone of the problem here is $\mathcal{K}=(k: k<-1)$, we
have one critical number $k=-2$.

The compatibility condition is
\begin{equation}2{C_0}^2-C_0\alpha\gamma-2\gamma^2=0\label{e:usl_sovm_r4}\end{equation}

If it holds, i.e.
\begin{equation}\alpha=2\frac{{C_0}^2-\gamma^2}{\gamma C_0}\label{e:usl_alpha_r4}\end{equation}
then, taking into account four terms, we obtain the expansion near
$z=\infty$
\begin{equation}\label{e:exp_r4}w(z)= \frac{C_0} {\gamma }\frac{1}{z}+
\frac{c_1}{z^2}+\frac{\gamma c_1^2}{C_0 z^3}+\frac{\gamma^2
c_1^3}{C_0^2 z^4}+\ldots\end{equation} where $c_1$ is the
arbitrary constant.

Denote
\begin{equation}P=2{C_0}^2-C_0\alpha\gamma-2\gamma^2\end{equation}
If compatibility condition \eqref{e:usl_sovm_r4} fails, i.e. $P\ne 0$,
the expansion is the power-logarithmic one. It takes the form
\begin{equation}\label{e:exp_r4_ln}
\begin{aligned}w(z)= &{\frac {{C_0}}{\gamma\,z}}+ \frac{c_1}{z^2}+\frac {P \ln z
}{{\gamma}^{3}z^2}  +\frac {\gamma\,{c_1}^{2}}{{C_0}z^3}-{\frac
{c_1\,P}{{\gamma}^{2}C_0z^3} }-{\frac {P \left( 3\,{\gamma}^{2}+2\,{{C_0}}^{2} \right)
}{{C_0}{\gamma}^{5}z^3} }+\\&+  \left( 2\,{\frac {c_1\,P}{{\gamma}^{2}{C_0}}}-{\frac
{{P}^{2}}{{C_0}{ \gamma}^{5}}} \right) \frac{\ln  z}{z^3} +{\frac {{P}^{2}
}{{C_0}{\gamma}^{5}}}\frac{\ln^2 z}{z^3}  +\ldots
\end{aligned}\end{equation}
where $c_1$ is the arbitrary constant.

\section{Expansions, corresponding to edge ${\Gamma_{5}}^{(1)}$.}

Edge ${\Gamma_{5}}^{(1)}$ exists, if $\beta=C_0=0$ and
$\gamma\ne0$.

The truncated equation, corresponding to edge
${\Gamma_{5}}^{(1)}$, takes the form
\begin{equation}\label{e:uk_ur_r5}w_{zz}+\gamma w^2=0\end{equation}

The outer normal $N_{5}=(1,-2)$, so the normal cone is
$$U_{5}^{(1)}=\{\lambda (1,-2), \lambda>0\}$$
Using the condition $\omega(1,r)\in U_{5}^{(1)}$, we obtain
$\omega=1$, $r=-2 \Rightarrow z\rightarrow\infty$

The solution of truncated equation \eqref{e:uk_ur_r5} can be
written as
\begin{equation}w=-\frac{6}{\gamma z^2}
\end{equation}
 For truncated equation \eqref{e:uk_ur_r5} we have
$$\mathcal{L}(z)=\dfrac{d^2}{d z^2}-\frac{12}{z^2}\quad \Rightarrow \nu(k)=k(k-1)-12$$

Equation $\nu(k)=0$ has roots $k=-3$ and $k=4$. As the cone of the
problem  is $\mathcal{K}=(k: k<-2)$, here there is one critical
number $k=-3$. The compatibility condition holds, if and only if
$\alpha=0$. In this case, taking into account three terms, we
obtain the expansion near $z=\infty$
\begin{equation}\label{e:exp_r5}
w(z)= -\frac{6}{\gamma z^2}+\frac{c_1}{z^3}-\frac {\gamma^{4}{c_{1}}^2+432}{8 \gamma^3 z^4}+\ldots\end{equation}
where $c_1$ is the arbitrary constant.

If $\alpha\ne0$, the expansion is the power-logarithmic one
\begin{equation}\label{e:exp_r5_ln}
\begin{aligned} w(z) =&-\,{\frac {6}{\gamma\,{z}^{2}}}+{\frac {72\,
\alpha\,\ln  z +7\,c_{{1}}{\gamma}^{2}}{{7\gamma}^{2}{z} ^{3}}}-\\-&{\frac
{648\,{\alpha}^{2} \ln^2  z-18\, \alpha\, \left( 24\,\alpha-7\,c_{{1}}{\gamma}^{2}
\right) \ln z
 }{{49\gamma}^{3}{z}^{4}}}+\\&+{\frac
{ 2160\,{\alpha}^{2}-21168+336\,\alpha\,{\gamma}^{2}c_{{1}}-
49\,{\gamma}^{4}{c_{{1}}}^{2} }{{392\gamma}^{3}{z}^{4}}}+\ldots
\end{aligned}\end{equation}
where $c_1$ is the arbitrary constant.
\section{Expansions, corresponding to edge  ${\Gamma_{6}}^{(1)}$.}\label{s:r6}

Edge ${\Gamma_{6}}^{(1)}$ exists, if $\beta\ne0$ or $\gamma\ne0$.

The truncated equation, corresponding to edge
${\Gamma_{6}}^{(1)}$, takes the form
\begin{equation}\label{e:uk_ur_r6}-2 w^3+\gamma w^2+\beta w =0\end{equation}

The outer normal $N_{6}=(1,0)$, so the normal cone is
$$U_{6}^{(1)}=\{\lambda (1,0), \lambda>0\}$$

Using the condition $\omega(1,r)\in U_{6}^{(1)}$, we obtain
$\omega=1$, $r=0 \Rightarrow z\rightarrow\infty$

The solution of truncated equation \eqref{e:uk_ur_r6} takes the
form
\begin{equation}\label{e:c_for_r6}w=c_0, \qquad
c_0=\frac{\gamma\pm\sqrt{\gamma^2+8\beta}}{4}
\end{equation}

Note, that solutions of truncated equation \eqref{e:uk_ur_r6} are
also the solutions of equation \eqref{e:eq}.

If solutions \eqref{e:c_for_r6} are the simple roots, i.e.
$\gamma^2+8\beta\ne0$, then there are no additional expansions.

If $\beta=-\gamma^2/8\ne0$, then $c_0=\gamma/4$ is the root of the second order. Then,
making substitution $w(z)=\gamma/4+v(z)$, for $v(z)$ we obtain equation \eqref{e:eq}
with coefficients $\widetilde{\beta}=0$, $\widetilde{\gamma}=-\gamma/2$ and
$\widetilde{C_0}=C_0-\alpha \gamma/4$. Coefficient $\alpha$ does not change. Let us
find expansions of function $v(z)$, which is the solution of equation \eqref{e:eq} with
modified coefficients, near $z=\infty$. Such expansions correspond either edge
${\Gamma_{4}}^{(1)}$ or edge ${\Gamma_{5}}^{(1)}$. Returning to initial notation, we
have the expansions for $w(z)$.

If $ \alpha=\dfrac{4{C_0}^2-\gamma^2}{\gamma C_0}$, $C_0\ne0$,
$\gamma\ne0$ or $\alpha=C_0=0$, we obtain the additional power
expansion near $z=\infty$. Otherwise, the additional expansion is
the power-logarithmic one.

%\section{Non-power asymptotic forms.}
%In sections 3-4 we found all the power asymptotic forms of
%solutions of equation \eqref{e:eq}. Here we are looking for their
%non-power asymptotic forms.
%
%According to \cite{Bruno02}, non-power asymptotic forms can be
%corresponded to edge ${\Gamma_{2}}^{(1)}$. This edge exists, if
%$C_0\ne0$ or $\beta\ne0$. The truncated equation, corresponding to
%this edge, takes the form
%\begin{equation}w_{zz}+C_0w_z+\beta w=0 \end{equation}
% Using the substitution
% \begin{equation}
%y=\frac{d\ln w}{dz}
% \end{equation}
%and taking into consideration
%$$w'=yw,\quad w''=(y'+y^2)w $$
%we obtain $$ w(y''+y^2+C_0 y+\beta )=0
%$$
%After cancellation of $w$ we have the equation
%\begin{equation}\label{e:as_f_eq}
%h(z,y)=y'+y^2+C_0 y+\beta
% \end{equation}
%The carrier of equation \eqref{e:as_f_eq} consists of points
%$(-1,1)$, $(0,2)$, $(0,1)$ (if $C_0\ne0$) and $(0,0)$ (if
%$\beta\ne0$).
%
%The cone of the problem here is $p_1=-p_2>0$. So we should consider
%only the vertical edge. The solutions of equation \eqref{e:as_f_eq},
%corresponding to the vertical edge, are
%\begin{equation}
%y=c_0,\quad c_0^2+C_0c_0+\beta=0
%\end{equation}
%
%Turning back to the initial notation, we obtain the non-power
%asymptotic forms of equation \eqref{e:eq}
%\begin{equation}
%w=c\exp{c_0 z}, \quad c_0^2+C_0c_0+\beta=0
%\end{equation}
%
%
%Note, that equation \eqref{e:as_f_eq} is Riccati equation, and its
%solutions are well-known.
\section{Exact solutions of equation \eqref{e:eq}.}

In sections 3- 9 we have obtained all the power asymptotic forms to
solutions of equation \eqref{e:eq}. All power and power-logarithmic
expansions, corresponding to these asymptotic forms, were also
found. They are the convergent ones \cite{Bruno01,Bruno02,Bruno03}.

Under some parameters the obtained power expansions can be summed,
and then we obtain the exact solutions of equation \eqref{e:eq}.

Let us give some examples of solutions, that can be found in such a
way.

Series \eqref{e:exp_r4} can be summed. So at $\beta=0$ and $\alpha$,
satisfying expression \eqref{e:usl_alpha_r4}, we obtain the exact
solution of equation \eqref{e:eq}
\begin{equation}\label{e:t_resh_r4} w(z)=\frac{{C_0}^2}{\gamma(C_0 z-c_1\gamma)}\end{equation}

The summation of expansion \eqref{e:exp_r5} (which exists if
$\alpha=\beta=C_0=0$) results in the solution of equation
\eqref{e:eq} in the form
\begin{equation}\label{e:t_resh_r5}
w(z) = -\frac{6/\gamma}{(z-z_0)(z-z_0-6/\gamma)}\end{equation} where
$z_0$ is the arbitrary constant, which can be expressed via $c_1$ as
$z_0=-(36+c_1\gamma^2)/(12\gamma)$.

Using the technique, described in section 9, we obtain the solution
\begin{equation}w(z)=\frac{\gamma}{4}-\frac{\gamma^2}{2C_0(\gamma z+2c_1C_0)}\end{equation}
on condition that
\begin{equation}\label{e:usl_1_r6}
\beta=-\gamma^2/8\ne0,\qquad \alpha=\frac{4{C_0}^2-\gamma^2}{\gamma
C_0}\end{equation} and solution
\begin{equation}w(z)=\frac{\gamma}{4}+\frac{12/\gamma}{(z-z_0)(z-z_0+12/\gamma)}\end{equation}
if
\begin{equation}\label{e:usl_2_r6}\beta=-\gamma^2/8\ne0,\qquad \alpha=C_0=0\end{equation}
Here $c_1$ and $z_0$ are the arbitrary constants.

\section{Conclusion.}
The Burgers-Huxley equation is used for description of some
nonlinear wave processes in physics, economics and ecology . In this
work we have studied the travelling-wave solutions of this equation,
using the power geometry methods.

We have found all power asymptotic forms and all corresponding
expansions of solutions of this equation at the different meanings
of parameters. We obtain power and power-logarithmic expansions:

1) power expansion \eqref{e:p_ex_v_1} near $z=0$ with two arbitrary constants

2) power expansion \eqref{e:p_ex_v_2} near $z=0$ with one arbitrary constant

3) two expansions near $z=0$, corresponding to asymptotic form \eqref{e:as_r1}, that can be power one \eqref{e:exp_r1}
as long as power-logarithmic one \eqref{e:exp_r1_ln} subject to the parameters of equation \eqref{e:eq}

4) two power-logarithmic expansions \eqref{e:as_r1} near $z=\infty$ (if $\beta=\gamma=0$ and $C_0\ne0$)

5) power expansion \eqref{e:exp_r4} near $z=\infty$, that can be summed and  gives the exact solution \eqref{e:t_resh_r4}
(if $\beta=0$, $\gamma\ne0$, $C_0\ne0$ and condition \eqref{e:usl_sovm_r4} holds)

6) power-logarithmic expansion \eqref{e:exp_r4_ln} near $z=\infty$
(if $\beta=0$, $\gamma\ne0$, $C_0\ne0$ and condition \eqref{e:usl_sovm_r4} fails)

7) power expansion \eqref{e:exp_r5} near $z=\infty$ with one
arbitrary constant, that can be summed and  gives the exact solution
\eqref{e:t_resh_r5}
 (if $\beta=C_0=\alpha=0$ and $\gamma\ne0$)

8) power-logarithmic expansion \eqref{e:exp_r5_ln} near $z=\infty$
 (if $\beta=C_0=0$ and $\gamma\ne0$, $\alpha\ne0$)

9) expansion near $z=\infty$, described in section \ref{s:r6} (if $\gamma^2+8\beta=0$),
that can be power one as long as power-logarithmic one.

%Summing one expansion, we found the exact solution of the
%Burgers-Huxley equation in travelling waves.

The obtained expansions can be used for testing codes at modelling
of the wave processes, which can be described by the
Burgers-Huxley equation.

\bigskip

This work was supported by the International Science and Technology
Center (project B1213).

\end{document}